\documentclass[trackchanges,twocolumn,tighten]{aastex62} 

\usepackage{amsmath}

\usepackage{longtable}
\usepackage{soul}
\usepackage{CJK}

\newcommand{\code}[1]{\texttt{#1}}

\newcommand{\mesa}{\code{MESA}}
\newcommand{\MESA}{\mesa}

\newcommand{\Athena}{\code{Athena++}}

\newcommand{\Msun}{M_\odot}
\newcommand{\Rsun}{R_\odot}
\newcommand{\Lsun}{L_\odot}

\newcommand{\Pgas}{P_\mathrm{gas}}
\newcommand{\Pturb}{P_\mathrm{turb}}

\newcommand{\Teff}{T_\mathrm{eff}}

\newcommand{\rphot}{R_\mathrm{ph}}

\newcommand{\vperp}{v_\perp}

\newcommand{\Menv}{M_\mathrm{env}}

\newcommand{\appropto}{\mathrel{\vcenter{
		\offinterlineskip\halign{\hfil$##$\cr
	\propto\cr\noalign{\kern2pt}\sim\cr\noalign{\kern-2pt}}}}}

\newcommand{\ltapprox}{\lesssim}

\newlength{\apjcolwidth}
\newlength{\figwidth}
\newlength{\doublewide}
\begin{document}
\begin{CJK*}{UTF8}{gbsn}
\title{Type IIb Supernova Progenitors in 3D: Variability and Episodic Mass Loss revealed by Radiation-Hydrodynamics Simulations}

\shorttitle{SNe-IIb envelopes in 3D}
\shortauthors{Goldberg et al.}

\author[0000-0003-1012-3031]{Jared A. Goldberg}
\affiliation{Center for Computational Astrophysics, Flatiron Institute, New York, NY 10010, USA}

\author[0000-0002-2624-3399]{Yan-Fei Jiang (姜燕飞)}
\affiliation{Center for Computational Astrophysics, Flatiron Institute, New York, NY 10010, USA}

\author[0000-0001-8038-6836]{Lars Bildsten}
\affiliation{Department of Physics, University of California, Santa Barbara, CA 93106, USA}
\affiliation{Kavli Institute for Theoretical Physics, University of California, Santa Barbara, CA 93106, USA}

\author[0000-0002-8171-8596]{Matteo Cantiello}
\affiliation{Center for Computational Astrophysics, Flatiron Institute, New York, NY 10010, USA}
\affiliation{Department of Astrophysical Sciences, Princeton University, Princeton, NJ 08544, USA}

\correspondingauthor{J. A. Goldberg}
\email{jgoldberg@flatironinstitute.org}

\begin{abstract}
We present the first 3D Radiation-Hydrodynamics simulations of partially-stripped ($M_\mathrm{core}\sim10M_\odot$, $M_\mathrm{env}\sim0.1-1M_\odot$) Yellow Supergiant ($L\sim10^5$, $\Teff\approx5000-8000$K) envelopes, constructed with \Athena. These envelope models represent the progenitors of Type IIb supernovae (SNe-IIb), which have lost a substantial fraction of their H-rich envelope before undergoing core-collapse. The luminosity-to-mass ratio is high in these extended envelopes, and convection is strongly driven by Hydrogen- and Helium opacity peaks. This surface convection, coupled with changes in the opacity, sustains large-amplitude low-azimuthal-order radial pulsations, creating order-of-magnitude variability in the stellar luminosity on a timescale of tens of days. If persistent prior to a SN-IIb, these variations could herald the upcoming explosion. 
Supersonic fluid motions across the outer layers of the star lead to both successful and failed mass ejection events, which shape the circumstellar environment and drive episodic mass loss ($\sim10^{-6}-10^{-5}M_\odot/$yr, in outbursts). The resulting 3D gas distribution in the outer atmosphere, responsible for early-time supernova shock-breakout and shock-cooling emission, shows orders-of-magnitude fluctuations in both space and time at any given radial location. This intrinsically complex halo of bound and unbound material complicates predictions for early SN-IIb lightcurves relative to spherically-symmetric models. 
However, it does provide a natural, self-consistent explanation for the presence and diversity of dense circumstellar material observed or inferred around pulsating evolved stars.
\end{abstract}

\keywords{
hydrodynamical simulations --- radiative transfer --- stars: massive --- supernovae: IIb}

\section{Introduction\label{sec:intro}}

Stripped-envelope massive stars have gained increasing attention in recent literature, as products of either binary stellar evolution \citep[e.g.][]{Eldridge2017,Drout2023,Gotberg2023,OGrady2024,Ludwig2025}, 
or enhanced mass-loss either pre- or post- Red-Supergiant (RSG) phase \citep[e.g.][]{Drout2009,Massey2021b,Humphreys2023,Cheng2024}. 
Depending on the mass remaining in the Hydrogen-rich envelope ($\Menv$), the effective temperature ($\Teff$) ranges from red ($\Teff\lesssim5000$K, $\Menv\gtrsim0.5M_\odot$), yellow ($8000\,\mathrm{K}\lesssim\Teff\lesssim5000$\,K, $0.5M_\odot\gtrsim{}\Menv\gtrsim0.05M_\odot$), 
or blue ($\Teff\gtrsim8000$\,K, $\Menv\lesssim0.05M_\odot$) \citep[see, e.g.,][]{Gilkis2025}. Of particular interest are the bright Yellow Supergiants (YSGs), where it is an important open challenge to identify partially-stripped YSGs in later stages of stellar evolution, compared to YSGs crossing the Hertzsprung Gap after core Hydrogen burning, on ``blue loops" during core Helium burning, or as the products of mergers with extra massive H-rich envelopes. 

Some stars are inferred to be partially-stripped only after their eventual explosions as Type IIb SNe \citep[e.g.][]{Aldering1994,Nomoto1995,Claeys2011,Georgy2012,Yoon2017,Sravan2019,Sravan2020,Laplace2020,Gilkis2022,Ercolino2024}, which tend to reveal yellow progenitors in pre-supernova imaging \citep[e.g.][]{Smartt2009, Smartt2015,Reguitti2025}. 
SNe-IIb require low but nonzero $\Menv\sim0.1-1\Msun$ in order to explain the presence of Balmer lines from some Hydrogen in the ejecta, but lack a plateau which would result from the recombination of a substantial amount of Hydrogen \citep[see, e.g.][]{Dessart2011a,Dessart2024}. 

While full SN-IIb lightcurves encode the total ejected mass and radioactive Nickel content \citep[e.g.][]{Shigeyama1994,Haynie2023}, rapid photometric observations reveal that many SNe-IIb also have a first luminosity peak within the first days after explosion \citep[e.g.][]{Woosley1994,Tartaglia2017}. Interpreted as shock-cooling emission of the low-mass ($\Menv\lesssim1M_\odot$) outer envelope \citep{Rabinak2011,Piro2017,Park2024}, modeling efforts presently find order-of-magnitude differences in the envelope mass and radius recovered from different models of the same event \citep[e.g.][]{Pellegrino2023,Farah2025}. These discrepancies arise primarily from differing assumptions about the density profile of the progenitor star’s diffuse outer hydrostatic envelope and its circumstellar environment \citep[e.g.][]{Piro2015,Piro2021,Sapir2017,Morag2023,Morag2025}.

Regardless of the physical mechanism for mass-loss which may lead to the fine-tuned stripping of the H-rich envelope required to achieve radii and $\Teff$ in the `Yellow' range, the stripped-YSG regime is also a challenge for detailed numerical modeling. Partially-stripped YSG envelope masses are 1-2 orders of magnitude lower than in their RSG counterparts, with correspondingly larger $L/M_\mathrm{env}$ \citep[e.g.][]{Gilkis2022}, 
which renders them more susceptible to pulsational instabilities \citep{Heger1997}. 
Likewise, the pressure scale height, $H$, is also large (of order the stellar radius $\rphot$), indicating that convection, with characteristic length scales $\ell\propto H$ \citep{Prandtl1925,BohmVitense1958}, 
will include large-scale cells spanning much of the stellar envelope. However, in this regime, the convection occurs at optical depths where radiation transport is able to carry
significant flux and impact the envelope dynamics, both in terms of where the diffusion speed equals the sound speed $\tau<c/c_{s}$ discussed by, e.g., \citet{Jiang2015}, and in terms of the balance between radiative and convective flux $\tau<(P_{r}\,c)/(P_\mathrm{total}\,v_c)$ \citep{Schultz2020,Schultz2023a,Schultz2023b,Goldberg2022a,Jermyn2022}. 
Pressure from radiation, gas, and the turbulent motion of the fluid thus all impact the stellar structure.
This necessitates careful treatment of Radiation Hydrodynamics in both the energy and momentum equations, which is now achievable at scale due to numerical advances \citep{Jiang2014,Jiang2021} and the increase in computational power. 

In this work, we present the first 3D Radiation-Hydrodynamics simulations of stripped-envelope Yellow Supergiants using the open-source radiation hydrodynamics instrument \Athena\ \citep{Stone2020,Jiang2021}. In \S\ref{sec:simsetup} we describe our numerical setup and in \S\ref{sec:timevariation} we describe the time evolution of the fluid properties in a convective steady-state.  In \S\ref{sec:structure} we describe the changing density and velocity structure throughout the star, and discuss implications for early supernova emission and angular momentum-driven transients during collapse into a black hole. In \S\ref{sec:asteroseismology} we describe the imprint of convection and pulsations on the observable stellar luminosity. In \S\ref{sec:massloss} we describe the induced episodic mass loss and the changing circumstellar halo of material. We conclude in \S\ref{sec:conclusions}. 

\section{3D Simulation Setup\label{sec:simsetup}}

We construct two simulations in \Athena\  \citep{Stone2020}, following same setup as used \citealt{Goldberg2022a}. We use spherical polar coordinates covering the domain $(r,\theta,\phi)\in[r_\mathrm{in},r_\mathrm{out}]\times[\pi/4,3\pi/4]\times[\phi_\mathrm{in},\phi_\mathrm{out}]$, with static mesh refinement in regions of interest. For the first simulation YSG1L4.7, $r_\mathrm{in}=28.1\Rsun$, $r_\mathrm{out}=1374.2\Rsun$, $\phi_\mathrm{in}=0$ and $\phi_\mathrm{out}=\pi$. At the root level, the radial direction is resolved using $N_r=160$ cells logarithmically spaced while the $\theta$ and $\phi$ directions are resolved using $N_{\theta}=64$ and $N_{\phi}=128$ cells uniformly. An additional level of mesh refinement is used for $r<450\Rsun$ to achieve the resolution $\delta{}r/r=\delta\theta=\delta\phi=1.23\%$. For the second run YSG1L5.1, the corresponding parameters are $r_\mathrm{in}=30\Rsun$, $r_\mathrm{out}=6436\Rsun$, $\phi_\mathrm{in}=0$, $\phi_\mathrm{out}=2\pi$, $N_r=112$, $N_{\theta}=32$ and $N_{\phi}=128$. One level of refinement is used for $450\Rsun<r<1000\Rsun$ to achieve resolution $\delta{}r/r=\delta\theta=\delta\phi=2.46\%$ in that region. Additional refinement is used to increase the resolution within $30\Rsun<r<450\Rsun$ by a factor of 2 (1.23\%) and a factor of 4 (0.62\%) for the region $r<70\Rsun$. 

\Athena\ solves the full set of radiation hydrodynamic equations by evolving the frequency integrated specific intensities over discrete angles based on the numerical scheme developed by \citet{Jiang2021}. The equations and methods are the same as used by \cite{Goldberg2022a}. 
We motivate the initial conditions for our YSG simulations from stellar models with partially-stripped H-rich envelopes using the 1D open-source stellar evolution software \MESA\ \citep{Paxton2011,Paxton2013,Paxton2015,Paxton2018,Paxton2019,Jermyn2023} r15140. 
Following \citet{Goldberg2022a}, we construct our motivating \MESA\ models starting with the \texttt{test\_suite} case \texttt{make\_pre\_ccsn\_IIp}, varying the initial mass and setting the mixing length $\alpha=3$ in the H-rich envelope. These models are non-rotating. We make one addition to this setup: the H-rich envelope is stripped down to various masses with enhanced winds after core He depletion, by modifying the parameter \texttt{\textquotesingle{}Dutch\_scaling\_factor\textquotesingle{}} beginning after the central He fraction $<10^{-2}$. 
This approximately mimics both enhanced RSG winds or late-stage (case C) mass transfer in a binary system.
We terminate the runs at core carbon depletion, and select models at core C depletion where $\Teff$ is in the YSG range (5000K-10000K) typical of SN-IIb progenitors \citep{Smartt2009}. 

\begin{figure*}
\centering
\includegraphics[width=\textwidth]{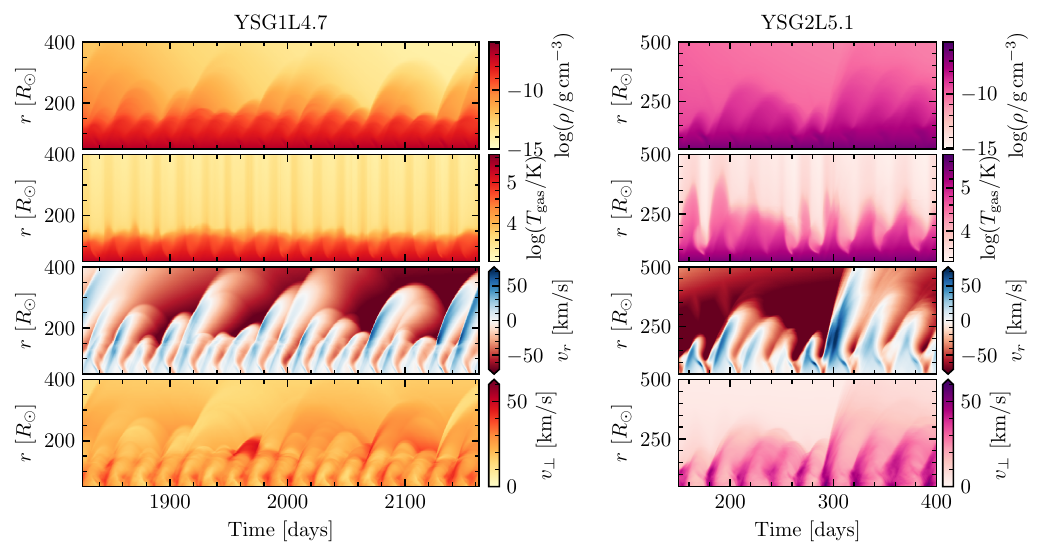}
  \caption{History of the angle-averaged radial profiles for the YSG1L4.7 (left) and YSG2L5.1 (right) models, zoomed in on a subset of times when the envelope is in convective steady-state. 
  Panels show  (top to bottom):  log$_\mathrm{10}$(density), log$_\mathrm{10}$(gas temperature), mass-averaged radial velocity, and magnitude of the mass-averaged tangential velocity.}
     \label{fig:spacetime}
\end{figure*}

We initialize the 3D model following the same procedure as in \citet{Goldberg2022a}. For the initial conditions we construct a hydrostatic radiative envelope with luminosity equal to the radiative luminosity at the radial location of the \Athena\ IB in the \MESA\ model ($R_\mathrm{IB}$). This is equivalent to $\approx$30\% of the total luminosity in the YSG regime, where convection is radiatively inefficient. The gas temperature at the IB is set to equal the temperature at the \MESA\ $R_\mathrm{IB}$ coordinate, and the density is then selected to approximately recover the total mass external to that location. To drive convection, the inner boundary is supplied with the larger total luminosity as radiation flux, via the ``Fixed L" boundary condition as in \citealt{Goldberg2022a}.
The models are labeled with the log of their time-averaged steady-state luminosity, with YSG1L4.7 indicating $\log(\langle{}L\rangle{}/\Lsun)=4.7$ and YSG2L5.1 indicating $\log(\langle{}L\rangle{}/\Lsun)=5.1$). 
The YSG1L4.7 (YSG2L5.1) model has $m_\mathrm{IB}=8.154\Msun$ ($11.77\Msun$) and a total mass of $8.2\Msun$ ($11.98\Msun$), with 0.018$\Msun$ (0.14$M_\odot$) directly in the simulation domain equivalent to $0.051\Msun$ (0.21$\Msun$) exterior to the IB. The motivating partially-stripped \MESA\ model had an initial mass of 21$\Msun$ (24$\Msun$) and He core mass of 7.1$\Msun$ ($10.6\Msun$). 
Stripped-envelope YSGs require stripping down into the H-burning shells in order to achieve the low envelope mass \citep{Long2022,Ercolino2024}; we use opacities for a mixture with $X=0.425$, $Y=0.555$, $Z=0.02$ and \citep{Asplund2009} abundances which match well the envelope abundances of our motivating \MESA\ models. We likewise adopt a mean molecular weight $\mu=0.78$. 
Both models have density floors imposed with $\rho_\mathrm{floor}=6\times 10^{-16}$g/cm$^3$.

YSG1L4.7 took approximately 6 months on 3200 cores on local high-performance computing resources, and the higher-resolution shorter-duration YSG2L5.1 run took $\approx$1 month on 6048 cores on NASA computing resources. 
We run until the models reach a steady-state for $r\gtrsim60R_\odot$ following the criteria outlined in \citet{Goldberg2022a} derived from the RHD equations (their Eq.~9): we determine when
$r^2\left\langle (E+\Pgas)\vr+F_{\mathrm{r},\hat{r}}-(\rho\vr\Phi)\right\rangle_t\equiv r^2\langle F_\mathrm{tot}\rangle$ is approximately spatially constant (within 5\%). 
In YSG1L4.7, this occurs after a few hundred days, and we analyze results after $800$ days. In YSG2L5.1, this occurs after $\approx$150 days. Due to the higher computational cost, we run this model for less time and primarily use it as a point of comparison, analyzing activity from 180 days onward.

\section{Time evolution of fluid properties \label{sec:timevariation}}

Fig.~\ref{fig:spacetime} shows the time-evolution of the shell-averaged density, temperature, and velocity during the convective steady-state. For clarity we show just a few hundred days of evolution, with the time since the simulation start given by the x-axis.
The shell-averaged $\rho$ and $T$ are calculated as the angle-average,
\begin{equation}
\langle{}X(r)\rangle\equiv\sum{X(r)\cdot\mathrm{d}\Omega}/\sum{\mathrm{d}\Omega},
\end{equation}
in each spherical shell at a given time, equivalent to a volume-weighted average at each radial location. Here $\mathrm{d}\Omega$ is shorthand for the solid angle area element $r^2\mathrm{d}(\cos\theta)\mathrm{d}\phi$ and $T_\mathrm{gas}$ is the gas temperature ($\Pgas/\rho$ in code units). A radiation temperature is calculated from $E_r$ as $T_r=(E_r/a_r)^{1/4}$, where $a_r=4\sigma_\mathrm{SB}/c$ is the radiation constant
and $\sigma_\mathrm{SB}$ is the Stefan-Boltzmann constant.
Although these simulations do not assume Local Thermodynamic Equilibrium, $T_r$ is typically within a few \% of $T_\mathrm{gas}$.
Shell-averaged radial velocity ($v_r$) is calculated as the mass-weighted of $v_r$ at each radius, $\sum\rho{}v_r\mathrm{d}r\mathrm{d}\Omega/\sum\rho{}\mathrm{d}r\mathrm{d}\Omega$, equivalent to the angle-average of the radial component of the momentum vector divided by the mass in each spherical shell. The average tangential velocity ($v_\perp$) is calculated shell-by-shell as the mass-weighted root-mean-squared (rms) tangential velocity $\sum\rho{}\sqrt{v^2_\theta+v_\phi^2}\mathrm{d}r\mathrm{d}\Omega/\sum\rho{}\mathrm{d}r\mathrm{d}\Omega$.
The tangential velocities are comparable to the radial velocity, fluctuating with the star's convective activity and pulsations. 

\begin{figure*}
\centering
\includegraphics[width=0.9\textwidth]{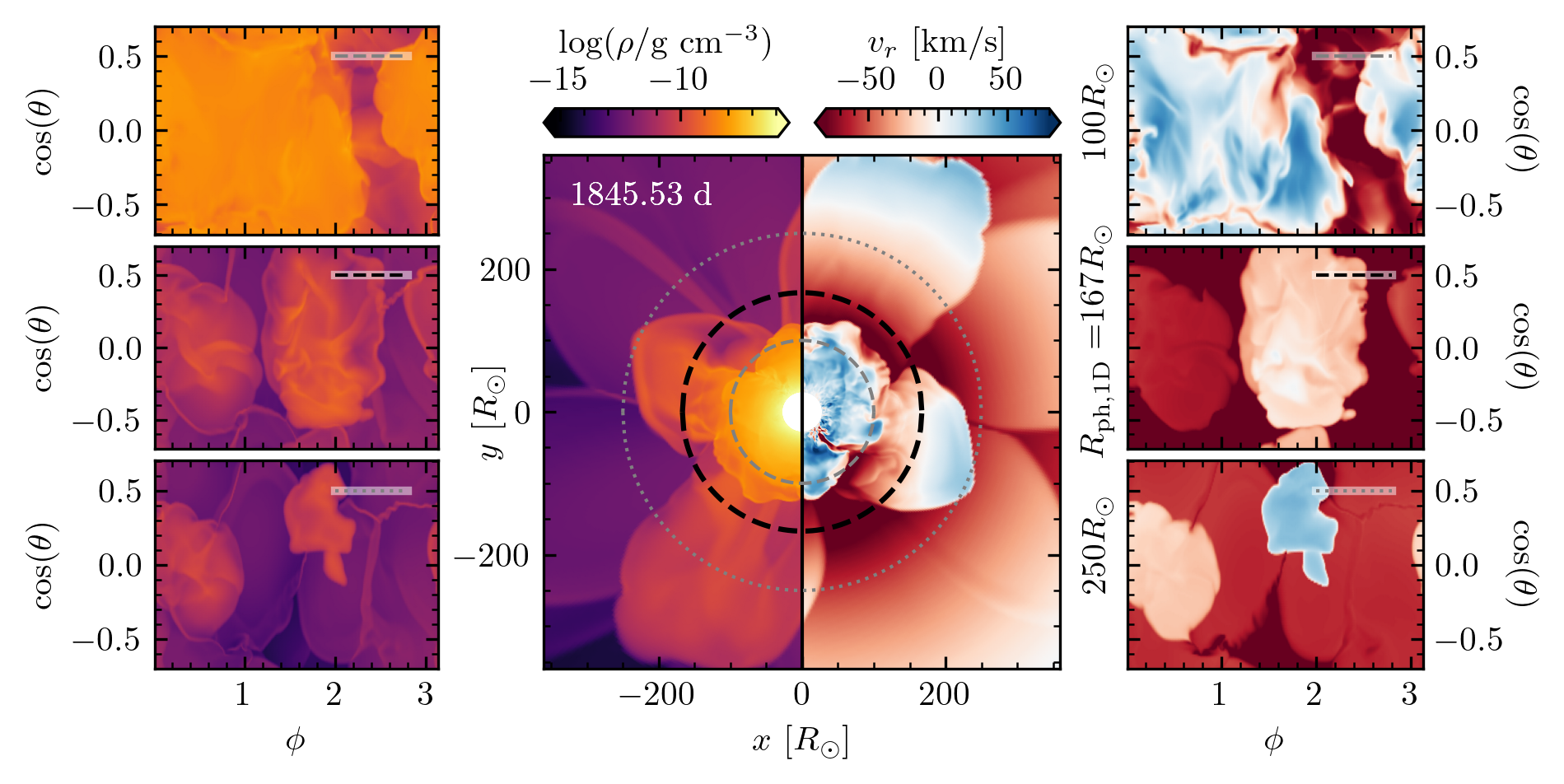}
\includegraphics[width=0.9\textwidth]{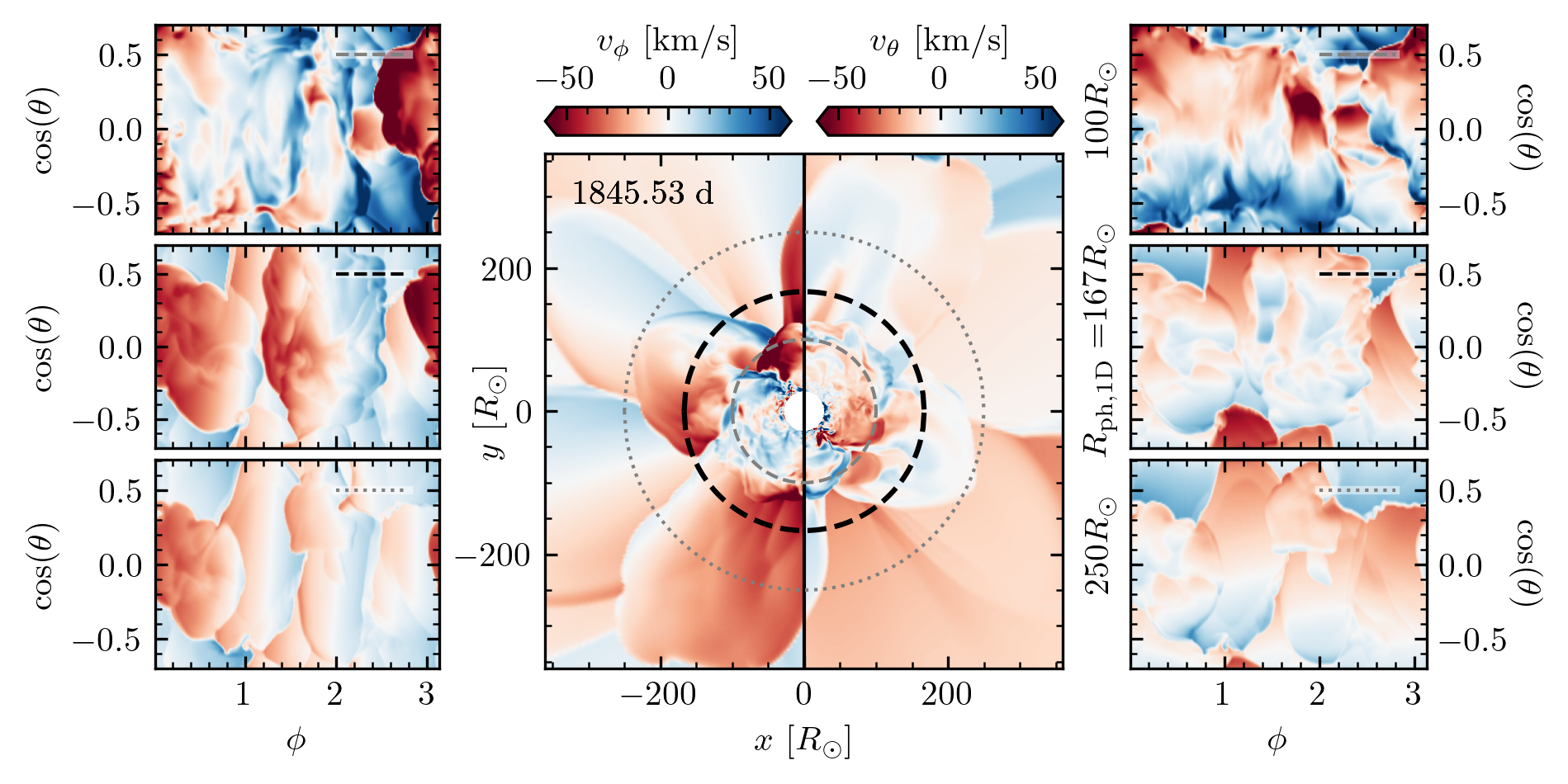}
  \caption{\textbf{Upper Panels:} Gas density (purple-orange colors, left panels) and radial velocity (red-blue colors, right panels) slices through a representative snapshot of the YSG1L4.7 model at day 1845.5. The central panels show equatorial slices, indicating fixed radii at $r=100R_\odot$ (grey dashed circle),  $r=R_\mathrm{ph}=167R_\odot$ (black dashed circle), and $r=250R_\odot$ (grey dotted circle). These radii, respectively, correspond to the upper, middle, and lower rows in the side columns, which show the variations in fluid properties across the simulation domain at fixed radius.  \textbf{Lower Panels:} Same as upper panels, but for azimuthal ($v_\phi$; red=counterclockwise) and co-polar ($v_\theta$; red=into-page) velocity. An animated version of the upper panel of this figure can be found \href{https://youtu.be/BjyGyG_PKns}{here}.}
     \label{fig:flower}
\end{figure*}

Calculating the radius and temperature of the `average' photosphere in 3D RHD models of giant star convection is notoriously difficult \citep[e.g.][]{Freytag2002,Chiavassa2011b,Chiavassa2024}, 
given large spatial-scale fluctuations in $T_\mathrm{gas}$ near the outer layers. For example, \citet{Chiavassa2011b} and \citet{Goldberg2022a} both adopt a photosphere where $L_\mathrm{surf}=4\pi{}R_\mathrm{phot}^2\sigma_\mathrm{SB}\langle{}T_\mathrm{gas}\rangle{}^4$, and equate the effective temperature $\Teff$ with the gas temperature at that location. However, when large-scale protrusions emerge from the stellar atmosphere, seen in the outbound lobes in the upper panels of Fig.~\ref{fig:flower}, this approximation can break down as the \textit{average} temperature at those radii does not represent the temperature of optically thick material. 
Alternatively, \citet{Jiang2015} and \citet{Goldberg2022b} identify the average radial location where $F_r/(cE_r)=1/3$ with the photosphere, which corresponds to where the specific intensity primarily points radially outward. 
This captures where the radiation flux thermodynamically decouples from the gas, so the radius where this happens better reflects the underlying radiation temperature an observer might determine from the spectrum. Therefore we choose this definition, and take the photosphere radius $\rphot$ as the angle-average of that radius. At most times, these two methods differ only by at most a few radial zones (a few $R_\odot$), although they can differ more when a dense plume protrudes well outside the stellar envelope, and when the stellar surface is undergoing significant compression. 

In both models, a pulsation persists through the steady-state. This pulsation is excited by convection, and sustained by changes in the surface opacity due to the temperature sensitivity of the hydrogen opacity peak and first helium peak.
For YSG1L4.7, $\rphot$ ranges from $\approx135-185R_\odot$. Even larger-amplitude photosphere fluctuations from 120-250$R_\odot$ are seen for YSG2L5.1. 
The changing temperature is evident in the time-lapse of $T_\mathrm{gas}$ in Fig.~\ref{fig:spacetime}, with dark colors at larger radii (higher $T_\mathrm{gas}$) corresponding to the hot phase of the pulsation, and lighter colors corresponding to the cooler phase. We define $\Teff\equiv{}(L/4\pi{}\sigma\rphot^2)^{1/4}$. The $\Teff$ of
YSG1L4.7 varies from $\approx$4800K to 9000K, and YSG2L5.1 ranges from $\approx$5000K to 10000K.
The pulsation periods remain steady, though there is stochasticity introduced due to the convective nature of the stellar envelope. We characterize the impact of these phenomena on the stellar luminosity in \S\ref{sec:asteroseismology}. 

During the pulsation cycle, material is elevated above the stellar photosphere out to large radii, up to a few times $\rphot$, leading to the orders-of-magnitude changes in density at large ($r>200R_\odot$) radii. Even when the outbound material is unbound, some material still falls back onto the stellar surface, with negative velocities seen in the third row panels in Fig.~\ref{fig:spacetime}. These successful and failed mass-ejection episodes impact the mass loss and structure of the circumstellar halo of material, which we quantify in \S\ref{sec:massloss}. 

\section{Envelope Structure \label{sec:structure}}
To illustrate the spatial variations throughout the stellar envelope, Fig.~\ref{fig:flower} shows equatorial and fixed-radius slices of a representative snapshot of the YSG1L4.7 model. The upper set of panels gives the density (upper left) and radial velocity (upper right), and the lower set of panels gives the azimuthal (lower left; red=counterclockwise) and co-polar (lower right; red=into-page) components of $v_\perp$. At the surface, these plumes are super-sonic. The black dashed curve in the central panels of Fig.~\ref{fig:flower} indicates $\rphot$. The central panels give equatorial slices, and side panels show the spatial distribution at fixed radius corresponding to the gray dashed, black dashed, and gray dotted lines shown in the central panels, at the radii labeled on the right panels. 
These models reveal a complex structure, with large-scale fluctuations from surface-layer convection and small-scale fluctuations from convection in the deeper interior. Around the star, outbound and infalling material enters and leaves the circumstellar halo. 

\begin{figure*}
\centering
\includegraphics[width=0.95\textwidth]{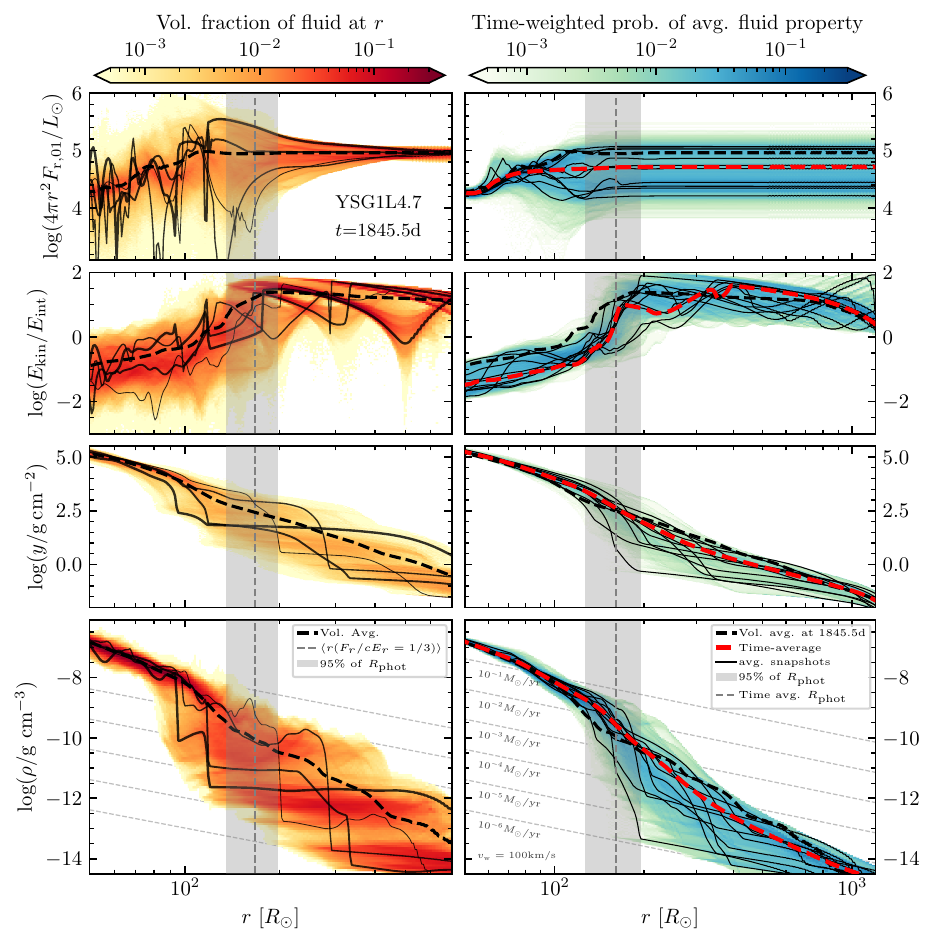}
\caption{\textbf{Left Panels:} Top to bottom show the line-of-sight luminosity $4\pi r^2 F_\mathrm{r,01}$, ratio of kinetic energy to internal energy, radially-integrated column density $y$, and gas density $\rho$ for a representative snapshot of the YSG1L4.7 model. Orange colors indicate the volume-weighted probability of finding a fluid element with a given y-axis value at each radial coordinate. Black curves of varying thickness show four arbitrary individual lines of sight through the envelope at fixed $\theta$ and $\phi$. Black dashed curves show the angle-averaged y-axis quantities. Vertical gray dashed lines show the average ``photosphere" location where $\langle{}F_r/cE_r\rangle=1/3$, with the gray shaded region indicating the radial span of 95\% of radii where the $F_r/cE_r=1/3$ condition is reached along different lines of sight. 
Thin gray dashed lines in the bottom row show wind-like density profiles with $\rho_\mathrm{w}(r)=\dot{M}/(4\pi{}r^2v_\mathrm{w})$, for $v_\mathrm{w}=100\mathrm{km/s}$ and varying $\dot{M}=10^{-(1-6)}\Msun/\mathrm{yr}$, with labels in the lower right panel. 
\textbf{Right panels:} The same quantities as a function of radial coordinate, but the blue-green colors indicate the time-weighted probability of finding the angle-average of each y-axis quantity at each radial coordinate at any given time. Thin black curves correspond to angle-averaged properties at individual snapshots. Black dashed curves are common to both columns and show the angle-average values in the fiducial snapshot. Dashed red curves in the right column indicate the time-average angle-average of all snapshots since the model has reached steady-state outside 50$R_\odot$. The vertical grey band shows the 95\% by-time range of $\rphot$, and the time-weighted average is indicated by the vertical grey dashed line.}
\label{fig:profiles}
\end{figure*}

\subsection{Spatial fluctuations and the shocked circumstellar halo}\label{sec:fluctuations} 

Fig.~\ref{fig:profiles} shows radial profiles of YSG1L4.7 varying in space and time. In the left panels, we show variations in fluid properties for a single snapshot, with thin solid lines marking individual random lines of sight and the colormap corresponding to the volume fraction of material with values indicated by the y-axis, at each radius. The average photosphere is shown by a vertical gray dashed line, with the gray shaded region indicating the interval where 95\% of the fluid reaches photospheric conditions $F_r/cE_r=1/3$. 
Dashed black lines in both the left and right panels indicate the volume-average of each quantity for the fiducial YSG1L4.7 snapshot. 

The right panels show instead the time evolution of the angle-averages, where thin lines correspond to the angle-averaged quantities at different times, and the colormap corresponds to the time-weighted rather than volume-weighted fraction of material for each quantity at each radius. The red dashed lines in the right panels give the time-averaged volume-average of each quantity during the steady-state portion of the simulation. Here the time-average angle-averaged photosphere is shown by a vertical gray dashed line, with the gray shaded region indicating the 95\% interval of $\rphot$ for different snapshots, which is comparable but not identical to the spatial spread within our fiducial snapshot. 

From the upper left panel of Fig.~\ref{fig:profiles}, which shows the luminosity profile from radiative diffusion along each line of sight ($4\pi{}r^2F_{r,01}$, left), we see that beneath the photosphere there are 2 orders of magnitude of diversity in the radiation flux along different lines of sight at a given radial location as large variations in the density and opacity provide channels for the radiation flux to diffuse. This converges to $\approx$20\% diversity even at larger radii well outside the photosphere. 
Likewise, deeper in (beneath $\approx120R_\odot$ in the fiducial snapshot), the integrated radiative flux does not account for the total luminosity leaving the star; the bulk fluid motion transports substantial energy flux outwards. 
The upper right panel of Fig.~\ref{fig:profiles} shows the total diffusive luminosity profile ($4\pi{}r^2\langle F_{r,01}\rangle$, right) at different times. The outgoing luminosity varies by an order of magnitude at different times, visible at large radii in the upper right panel. The time-averaged luminosity (red) shows that convection dominates the energy transport only out to $\approx80R_\odot$, even though fluid motion and corresponding inhomogeneities persist to well outside the photosphere.

The second row gives the ratio of kinetic energy to internal energy, proportional to the Mach number squared ($\mathcal{M}^2$). 
In isotropic convection, the turbulent pressure $\Pturb$ is related to the kinetic energy density $e=\rho v^2/2$ by $\Pturb=2e/3$, so this similarly gives a measure of the turbulent pressure relative to the thermal pressure. Within $\approx100\Rsun$, the average fluid motion remains sub-sonic at all times, but still with $\mathcal{M}\sim0.5$. Moreover, within a given snapshot, some fluid reaches sonic motion even in the deep interior. Outside the photosphere, the kinetic energy exceeds the thermal energy by a factor of 10-100 ($\mathcal{M}\sim3-10$), with shocks forming between the infalling material and the outgoing (attempted) mass ejections. 
These shocks can be seen as sharp transitions along individual lines of sight in the third and fourth panels, which show the column depth $y=\int_r^\infty{\rho(r)\,\mathrm{d}r}$, and the density $\rho(r)$, respectively. 

\subsection{Implications for SN-IIb shock breakout and cooling}
The stellar density profile is crucial for interpreting the supernova shock breakout and shock cooling emission. If fully ionized, e.g. during the supernova explosion as the shock approaches shock breakout, the column depth is related to the optical depth by the electron-scattering opacity $\tau_s=\kappa_sy$, and the shock will break out of the ejecta at $\tau_s=c/v_\mathrm{sh}$, or $y=c/(\kappa_sv_\mathrm{sh})$ where $v_\mathrm{sh}$ is the shock velocity. Because the material is He-enriched relative to solar abundances, the scattering opacity is $0.29$~cm$^2$\,g$^{-1}$, so shock breakout will occur around $y$ of $\sim$30-300~g/cm$^2$ ($\log(y/\mathrm{g\,cm^{-2}})\sim1.5-2.5$) depending on the shock velocity. 
If the SN occurs instantaneously during our fiducial snapshot, this will lead to a very oblique shock breakout along outgoing plumes. At a given time, the radial span of the potential breakout region is greater than the photospheric radius of the star itself, this will lead to a large timing difference between shock breakout in the innermost and outermost regions of the optically thick stellar wind. This timing difference could smear out the breakout signal \citep{Goldberg2022b} in addition to the smearing due to oblique breakout along the asymmetric lobes \citep[e.g.][]{Matzner2013,Salbi2014,Irwin2021,Irwin2025}. 

Moreover, the density itself varies by 4 orders of magnitude at near-photospheric radii, both for the individual snapshots and for the angle-averages considered at different times. 
Thin gray dashed lines in the bottom row show $1/r^2$ wind-like density profiles for varying constant $\dot{M}=10^{-(1\dots6)}\Msun/\mathrm{yr}$ and fixed wind velocity $v_\mathrm{w}=100\mathrm{km/s}$, with $\rho_\mathrm{w}(r)=\dot{M}/(4\pi{}r^2v_\mathrm{w})$. If we were to interpret the material out at 1000$\Rsun$ as an outgoing wind assuming spherical symmetry, we would recover different $\dot{M}$ by over an order of magnitude at different times. Even greater discrepancies arise as we probe bound material closer to the stellar surface, with factor-of-1000 variation in $\langle\rho\rangle$ and thereby an inferred $\dot{M}$ at $\approx2\rphot$. Moreover, the slope of the density profile, which is crucial for shock cooling modeling, also varies dramatically along each line of sight, with strong discontinuities caused by the supersonic fluid motion as well as the contact shock which forms between the infalling and outbound material (lower left panel of Fig.~\ref{fig:profiles} and upper left panels of Fig.~\ref{fig:flower}). Likewise, the average density profile changes a function of time, primarily near the photosphere and in the surrounding circumstellar halo (lower right panel of Fig.~\ref{fig:profiles}). 

We emphasize that this large diversity of outer density profiles are achieved in the same star, shaped by the pulsation phase, convective plumes, and mass-loss episodes which we further discuss in \S\ref{sec:massloss}. 
As the eventual supernova shock will take less than a day to reach the stellar surface, but the envelope and surrounding material change on a tens-of-days-timescale, explosions at different times will see vastly different density structures.
Shock breakout and the subsequent cooling emission are particularly sensitive to the outer density profile \citep{Rabinak2011,Piro2021,Khatami2024} in the low-envelope-mass regime; thus the varying physical structure directly impacts inferences of ejecta properties from early SN-IIb lightcurves for the first few days, when the supernova photosphere lives within this outer envelope. 

\subsection{Stochastic Angular Momentum}
The stochastic angular momentum (AM) from convection plays an important role in black hole (BH) formation as a supergiant star with a convective envelope attempts to collapse \citep{Coughlin2018,Quataert2019,Antoni2022}. 
At every radial location, if the convection contains large-scale fluctuations, then the specific angular momentum in a thin shell will be nonzero. As the star collapses, these layers attempt to accrete into the BH over time.
If the shell-by-shell specific angular momentum $j_\mathrm{rand}$ gives a circularization radius $r_\mathrm{circ}=j^2/GM$ which is greater than the innermost stable circular orbit of the BH, $r_\mathrm{isco}=6GM/c^2$ in the case of Schwarzschild BH, then a disk will form, and the infalling material will need to transport its angular momentum before falling into the BH. In this case, as more and more layers begin to circularize, a weak shock forms which launches a low-luminosity, long-duration transient that can eject the majority of a H-rich envelope \citep{Antoni2023,De2024} from the forming BH. 

As seen in the lower panels of Fig.~\ref{fig:flower}, there are substantial tangential velocities despite the models being non-rotating. 
In the deeper interior ($r\ltapprox100R_\odot$), the convection is smaller-scale compared to RSGs, as $H/r\approx0.1$ in our models compared to $H/r\approx0.25$ in RSGs \citep{Antoni2022, Goldberg2022a, Ma2024, Ma2025}. Nonetheless, asymmetries in $\vperp$ lead to nonzero cancellation of the total angular momentum in each shell ($|j_\mathrm{rand}|\sim10^{18}$cm$^2$/s), despite the total angular momentum being zero integrated through the whole simulation domain. 
This value is slightly smaller than in RSGs ($|j_\mathrm{rand}|\sim\mathrm{a\ few\times}10^{18}$cm$^2$/s) constructed with a similar numerical setup \citep{Goldberg2022a}, and comparable to scale-free pure hydrodynamic simulations for this mass regime \citep{Antoni2022}. 
In the outer layers, however, the non-purely-radial pulsations lead to a steep increase in $||j_\mathrm{rand}||$ near the photosphere, exceeding $10^\mathrm{19}$cm$^2$/s in the outer regions above $~150R_\odot$ in both YSG1L4.7 and YSG2L5.1.
This entails circularization radii of a few$\times10^{-2}R_\odot$ in the convective interior, and $\sim1-10R_\odot$ at the surface, in both cases much larger than the $r_\mathrm{isco}=1.03\times10^{-4}R_\odot$ and $1.49\times10^{-4}R_\odot$ for YSG1L4.7 and YSG2L4.9, respectively. We therefore do expect circularization of the infalling fluid during the birth of a BH embedded inside a partially-stripped envelope such as the ones presented here. 

Moreover, the mass in the envelope is 1-2 orders of magnitude lower, $\sim0.1-1\Msun$ compared to $\sim10\Msun$. Thus, a stochastic-AM jet-driven explosion engine has 2-3 orders of magnitude \textit{less} energy to tap into (assuming collapse to an equivalent BH mass). We thus expect stripped-envelope YSG collapse to, in general, lead to much \textit{lower}-luminosity transients from BH birth as compared to RSG envelope collapse, consistent with the discussions in \citet{De2024}. 
We likewise note that all of the direct-collapse BH candidates \citep{Kochanek2024,Beasor2024,De2024}, where no luminous transient was observed, come from progenitors with yellow $\Teff$ ($\approx4500-6000$K) at some point before they disappeared from optical bands.

\begin{figure}
\centering
\includegraphics[width=\columnwidth]{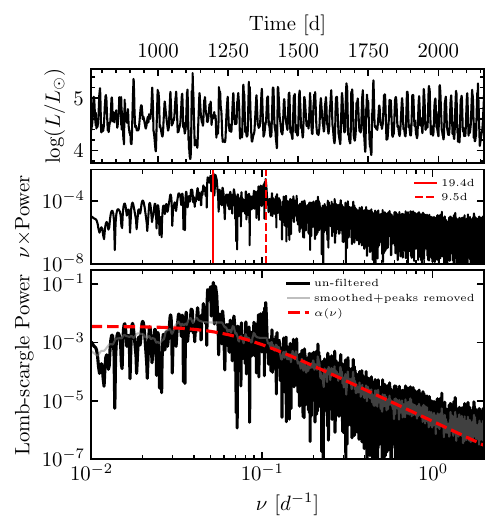}
  \caption{\textbf{Upper~panel:} Light-curve of YSG1L4.9. \textbf{Middle~panel:} Frequency-weighted power spectrum, with vertical red lines indicating the two  pulsation modes at 19.4d (solid) and 9.5d (dashed). \textbf{Lower~panel:} Lomb-scargle power spectrum (solid black) as well as the smoothed power spectrum with peaks removed. The fitted SLF variability function (Eq.~\ref{eq:slf}) is shown as a red dashed line.}
     \label{fig:variability}
\end{figure}

\section{Stellar variability\label{sec:asteroseismology}}
We construct lightcurves by calculating the integrated radiation flux $\int{}F_{r,0}\mathrm{d}A$ leaving the simulation domain, normalized to the solid angle of the simulation domain, as in \citet{Goldberg2022a}. These are semi-global models encompassing $\approx$70\% of the face-on hemisphere (and 70\% of the full 4$\pi$, in the case of YSG2L5.1); the variations from $\sqrt{N}$ statistics from convection including the poles will be small. Both models exhibit periodic and stochastic variability in the lightcurves. 

Figure \ref{fig:variability} shows the lightcurve and power spectra for the YSG1L4.7 model, which shows periodic and stochastic luminosity fluctuations ranging from $\log(L/L_\odot)\approx4-5.4$. 
Due to uneven time-sampling of our simulation output, the power spectra were calculated with a Lomb-Scargle periodogram \citep{Lomb1976,Scargle1982,Townsend2010} using the Python framework SciPy \citep{SciPy}. 
Two peaks in the YSG1L4.7 power spectrum are present at 19.4 days and 9.5 days, indicated by vertical lines in the middle panel. These are p-modes in the outer envelope, with low radial and azimuthal order ($\ell=2$) sustained by convective excitation and as opacity changes. 
Variability of this nature to be expected due to the high $L$ and low $\Menv$ compared to their RSG counterparts; larger $L/M_\mathrm{env}$ entails stronger susceptibility to pulsational instability \citep{Gough1965,Heger1997,Aerts2009,Yoon2010}. 
The large pulsation amplitude accounts for $\approx2/3$ of the root-mean-squared variance in the lightcurve. 
Similarly, the YSG2L5.1 model has one identifiable mode at 35 days, with luminosity fluctuations ranging from $\log(L/L_\odot)\approx4.5-5.6$. 
These differences between the two models are broadly consistent with brighter luminosities entailing longer-period pulsations, though this should not be interpreted as a true Period-Luminosity relation given the differences in the geometry of the simulation domain. 

The additional $\approx1/3$ of the total power comes from the stochastic variability. 
Though ubiquitous, the cause of stochastic low-frequency (SLF) variability in luminous star observations remains debated. In OB stars, it was initially attributed to a spectrum of gravity \mbox{(g-)} modes from the core propagating through the stellar envelope \citep[e.g.][]{Bowman2019a,Bowman2019b}, with others arguing that it comes from the subsurface convection driven at the Fe and He opacity peaks \citep{Cantiello2021,Schultz2022,Schultz2023b,Anders2023,Pedersen2025}. 
In our models, the SLF variability comes from the turbulent fluid motion in the convective envelope; there is no core within the simulation domain from which g-modes could be driven. 

\begin{figure*}
\centering
\includegraphics[width=0.95\textwidth]{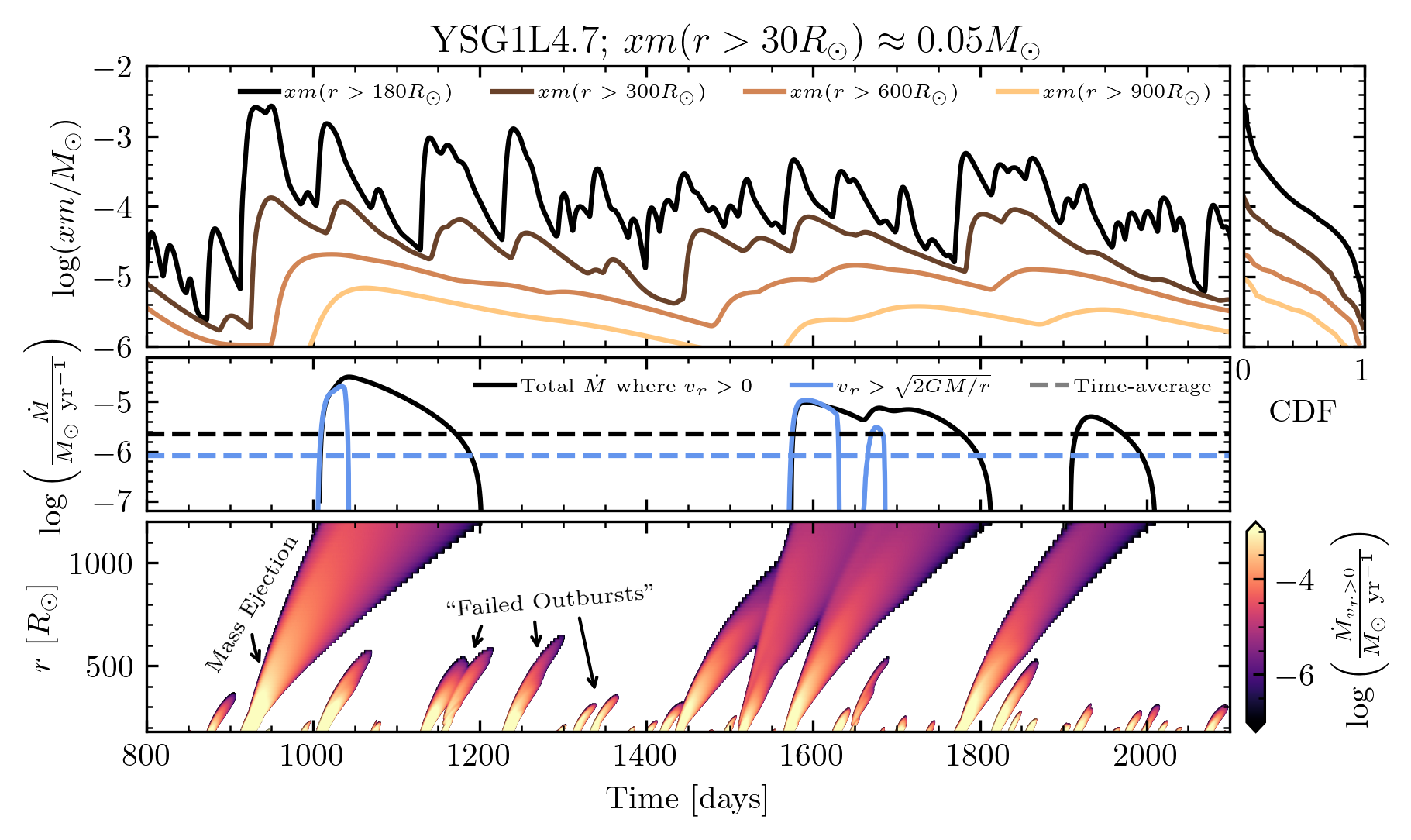}

\caption{Eruptive Mass loss and the resulting circumstellar halo for the YSG1L4.7 model. \textbf{Upper left panel:} Mass exterior to select radii  ($xm$) as a function of time. The black line at $r=180R_\odot$ approximates the mass external to the photosphere. \textbf{Upper right panel:} X-axis shows the Cumulative Distribution Function (CDF) of each $xm(r,t)$ time-series  (y-axis) binned and integrated from high-$xm$ to low-$xm$, equivalent to the fraction of time spent with $xm\ge$ the y-axis value. \textbf{Middle panel: } Integrated mass flux through the $1200R_\odot$ location, giving upper and lower bounds on the unbound mass: the total outward mass flux (black, upper bound) and the mass flux with $v_r>\sqrt{2GM/r}$ (blue, lower bound). Time-averaged values are indicated by horizontal dashed lines. \textbf{Lower panel:} Outward ($v_r>0$) mass loss rate as a function of time (x-axis) and radial location in the simulation (y-axis); lighter colors indicate larger $\dot{M}$.}
     \label{fig:eruptions}
\end{figure*}

To quantify the SLF variability, we follow the cleaning and fitting procedure of \citet{Schultz2022}, wherein the peak values are reduced to the mean of the neighboring frequency bins in the power spectrum and a simple moving average is used to smooth out the noise in the power spectrum, the result of which is represented by the faint gray curve in the lower panel of Fig.~\ref{fig:variability}.  We then fit the cleaned power spectrum using nonlinear least squares to the function \citep{Bowman2019a,Bowman2019b,Schultz2022,Pedersen2025}: 
\begin{equation}
\alpha(\nu)=\frac{\alpha_0}{1+\left(\frac{\nu}{\nu_{\mathrm{char}}}\right)^\gamma},
\label{eq:slf}
\end{equation}
where $\alpha_0$ is the characteristic amplitude, $\nu_{\mathrm{char}}$ is the characteristic frequency over which the turnover happens, and $\gamma$ is the power-law decay at high frequencies.
This fit is given by the red dashed curve on the lower panel of Fig.~\ref{fig:variability}. We recover $\nu_\mathrm{char}=0.065\,\mathrm{d}^{-1}$ ($0.029\,\mathrm{d}^{-1}$) and $\gamma=-2.7$ ($-2.6$) for YSG1L4.7 (YSG2L5.1).

Importantly, even from the SLF ``red noise" fit which falls off at higher frequencies, we continue to see power at the $1\%$ level out to shorter timescales of a few days (frequencies of $\approx0.3\,\mathrm{d}^{-1}$). The envelopes are highly convective, therefore some variability on timescales shorter than the pulsation period will naturally arise from the turbulent cascade. We additionally ran with highly time-sampled output (0.01d) for $100$d to test for very high-frequency variability. We do not see any evidence in either model for periodic signals on such short timescales to be consistent with the Fast Yellow Pulsations (FYPS) proposed for some more luminous $L\gtrsim10^5L_\odot$ YSGs (\citealt{DornWallenstein2020,Dorn-Wallenstein2022}). In many cases, the faster-than-acoustic-cutoff pulsations have been demonstrated to be contamination from nearby tight binaries in the TESS data \citep{Pedersen2023}.

\section{Eruptive mass loss and the changing circumstellar envelope \label{sec:massloss}}

Due to the presence and motion of material elevated in the pulsations, there are substantial variations in the mass within the halo of material outside the photosphere. Some of this material is unbound, and escapes the simulation domain, whereas other material falls back onto the stellar surface, in some cases after reaching up to $2-3$ times the stellar radius. 
Fig.~\ref{fig:eruptions} illustrates the connection between the episodic unbound mass loss, failed mass ejections, and changing circumstellar halo. 
The lower panel shows a space-time diagram of the local mass loss rate calculated from the outward mass flux, $\dot{M}=4\pi{}r^2\langle\rho{}v_r\rangle$ where the average is taken only over the zones where $v_r>0$. White regions in the lower panel indicate purely inward motion at that radius at that time. The central panel shows the instantaneous mass loss rate assuming all outward-moving material at $r=1200R_\odot$ escapes to $\infty$ (black), as well as the most conservative estimate of the unbound mass loss rate (blue), including only zones where $v_r>v_\mathrm{esc,grav}=\sqrt{2GM/r}$. 
Horizontal dashed lines show the time-averaged mass-loss rates for the two conditions, which for the YSG1L4.7 model gives conservative lower and upper bounds of $\approx10^{-6}M_\odot/$yr ($v_r>v_\mathrm{esc}$) and $\approx3\times10^{-6}M_\odot/$yr ($v_r>0$). 
Though we have a shorter runtime for YSG2L5.1, we can still estimate a larger time-averaged $\dot{M}\sim10^{-5}\Msun\,\mathrm{yr^{-1}}$, with two separate mass ejection episodes making it out past $650\Rsun$ within 200 days. 
In both models, while the instantaneous average mass-loss changes with distance as ejected material expands, the time-average is consistent at large $r\gtrsim1000\Rsun$.

We also calculated the unbound mass loss rate where the gravity is modified to reflect additional support from radiation and the unbound mass is defined as outward-moving material with positive total enthalpy (not shown in Fig.~\ref{fig:eruptions}).
However, the stellar luminosity changes by an order of magnitude on a $\approx20$d period to nearly-Eddington luminosities. This is akin to turning on and off gravity with a flickering lightbulb under a marginally-unbound flow on a timescale shorter than the dynamical fallback time of material at large radii. 
Thus, most mass which makes it far away from the star's surface is indeed unbound; thus the mass loss rate is much closer to the total $\dot{M}$ than the naive estimate considering only the escape velocity compared to $v_r$. 
We also note that there is no dust in these simulations, whereas sufficiently far from the stellar surface ($\gtrsim5\rphot\sim1000\Rsun$), gas could become cool enough to begin to form dust. 
The presence of dust would serve to further levitate the outbound material, and bring the unbound mass loss rate closer to the total outward mass loss rate shown here \citep[see, e.g., discussions in][]{Fuller2024}. 

Both successful and failed `outbursts' create the star's surrounding material.
The changing mass in the outer halo is shown in the upper panel of Fig.~\ref{fig:eruptions}, which shows the exterior mass ($xm$) outside various radii as a function of time. Material just outside the photosphere ($\rphot\approx180R_\odot$) fluctuates from $<10^{-5}M_\odot$ to a few $\times 10^{-3}M_\odot$, as material gets launched into the outer halo, and either continues to leave, or falls back onto the stellar surface. 
Farther out, the periodic fluctuations diminish, as more material that leaves is likely to remain unbound and moving outward. 
Temporal changes correspond to the launching episodes, shown in the lower panel. 
In many cases, material reaches nearly 2 stellar radii ($\approx300-400R_\odot$) before falling back onto the star, evident in the dark brown curve corresponding to $300R_\odot$. 

This fallback can have observational consequences; infalling gas along the line of sight is expected to produce inverse P-Cygni profiles in spectroscopic data. In YSGs undergoing episodic mass loss, the competition between outflow and fallback could then produce time-varying P-Cygni and inverse P-Cygni morphologies. 
Such behavior has indeed been reported in some Galactic yellow supergiants and hypergiants, with inverse P-Cygni profiles interpreted as signatures of localized fallback close to the stellar surface. The infall velocities inferred for the material are typically on the order of tens of km s$^{-1}$, consistent with our simulations  \citep{Oudmaijer1998,Humphreys2002,Lobel2005,Klochkova2018}. 
Even in cases where there is insufficient outer material to contribute to emission, we should still expect varying degrees of line asymmetry in absorption features from the large spatial (Fig.~\ref{fig:flower}) and temporal (Fig.~\ref{fig:spacetime}) variations in the velocity field. This has recently been observed at the population level in single-epoch spectroscopic observations of YSGs \citep{Chen2024,DornWallenstein2025}.

We additionally show the amount of time which the star spends with some amount of material outside of each radius. The x-axis of the upper right panel of Fig.~\ref{fig:eruptions} shows the fraction of time the model spends with $xm$ greater than the same y-axis values as the upper-left panel. This is quantified by taking the cumulative distribution function (CDF) calculated from the ordered histogram of $xm$ values binned in time, integrated from high-mass to low-mass along the y-axis. For example, the YSG1L4.7 model spends no time with $xm>10^{-2.5}M_\odot$ beyond $r=180R_\odot$, 100\% of the time with  $xm>10^{-5.6}M_\odot$ beyond $r=180R_\odot$, and about half its time with $\gtrsim10^{-4}\Msun$ beyond $r=180R_\odot$. 

This is a useful guide for our expectations for the presence of `dense circumstellar material' out to a few stellar radii, often inferred in Supernova observations, discussed frequently in the context of SNe-IIP from Red Supergiants \citep[e.g.][]{Morozova2017,Morozova2017b,Irani2024,Jacobson-Galan2025,Das2025}, but also SNe-IIb's \citep[e.g.][]{Chevalier2010,Ouchi2017}. 
If the star explodes with no knowledge of the pulsation or mass loss phase, we would predict the fraction of IIb's from Yellow Supergiants with `dense' CSM to track the fraction of time the star spends with suspended material due to the pulsational eruptions, with $\sim$half of events showing evidence of $\sim1\%$ of the outer envelope mass present beyond the photosphere radius. 

\section{Discussion \& Conclusions\label{sec:conclusions}}

In this work, we present 3D Radiation-Hydrodynamics simulations of luminous low-mass H-rich envelopes in \Athena, and predict their observable properties prior to explosion as Type IIb Supernovae. These simulations reveal highly variable envelopes, with convection accounting for substantial energy transport in the stellar interior, shocks from pulsations and infalling material forming near the photosphere, and a circumstellar halo of material shaped by bound and unbound plumes which are ejected from the star. 
This manifests in order-of-magnitude variability in the stars' luminosity on a timescale of tens of days, with both steady pulsations and stochastic variability at low frequencies, which can be observable in time-domain surveys. The episodic mass ejections lead to modest total mass loss rates of $\sim10^{-6}-10^{-5}M_\odot/$yr, but load the circumstellar halo with an average of $\sim10^{-4}M_\odot$, with order-of-magnitude variance in time. 
At any given time, this could resemble a changing mass loss rate just prior to the explosion, which is inferred from observations of the canonical Type IIb event SN-1993J \citep[e.g.][]{vanDyk1994}. 

This highly-variable partially-stripped yellow envelope phase will persist until the stellar envelope structure is significantly impacted by the mass loss, or until the star explodes, whichever comes first.
After core C ignition, a massive star on its way to core-collapse has of order a few thousand years left to live \citep{Woosley2002}, which at $\dot{M}\sim3\times10^{-6}\Msun/\mathrm{yr}$ would correspond to a total mass loss of $\sim10^{-2}M_\odot$, a $\sim1-10\%$ correction to the envelope mass. Even if this phase persists beginning during core He burning, it would optimistically last on the order of $\sim10^4$ years. 
Recent work \citep{Das2023,Subrayan2025} gives a local rate of SNe-IIb of $\sim 5\times 10^{-6}\ \mathrm{SNe\ per\ Mpc}^3$ per year.
Hence, within a volume of 10Mpc, we expect about 1 SN-IIb every 50 years. Within that volume, if the phase of high progenitor variability is $\sim10^4$ years, then we expect about 200 active IIb progenitors (or tens to $\sim100$ if we consider the core C burning lifetime). 
The Vera Rubin Observatory's Legacy Survey of Space and Time (LSST; \citealt{Ivezic2019}) will see out to a limiting magnitude of $\approx24-27.5$ in $r$ band, which translates to a volume of $\gtrsim$10Mpc for resolving changes on the order of $10^4\Lsun$. Deep monitoring over the decade-long survey will therefore yield insight into the dynamic lives of the progenitors of these transitional events, and if we're lucky, direct observations of the progenitor variability prior to an observed explosion.\\


\acknowledgements
We are grateful for helpful conversations with Andrea Antoni, Shelley Cheng, Ayanna Mann, Brian Metzger, Mathieu Renzo, William C. Schultz, Lieke van Son, and Samantha Wu. We also thank Conny Aerts, Kishalay De, Trever Dorn-Wallenstein, Maria Drout, Joseph Farah, and Anna O'Grady for valuable discussions about observations of these systems.
This research benefited from interactions with a variety of researchers that were funded by the Gordon and Betty Moore Foundation through Grant GBMF5076.
The Flatiron Institute is supported by the Simons Foundation. 
This research was supported in part by grant NSF PHY-2309135 to the Kavli Institute for Theoretical Physics (KITP)
and by grant ATP-80NSSC22K0725 through the NASA Astrophysics Theory Program.
Computational resources were provided by the Flatiron Institute and the NASA High-End Computing (HEC) program through the NASA Advanced Supercomputing (NAS) Division at Ames.

\bibliographystyle{aasjournal}
\singlespace

\bibliography{RSG3D.bib}


\end{CJK*}
\end{document}